\documentclass[12pt,twoside]{article}

\usepackage[T2A]{fontenc}
\usepackage[utf8]{inputenc}
\usepackage[english]{babel}
\usepackage{amsmath}
\usepackage{amssymb}
\usepackage{aas_macros}
\usepackage{graphicx}
\usepackage{pazhutf}
\usepackage[tmargin=15mm, bmargin=30mm, lmargin=20mm, rmargin=20mm, footskip=0mm]{geometry}

\usepackage{xcolor}
\usepackage{hyperref}
\hypersetup{%
  colorlinks=false,
  pdfborderstyle={/S/U/W 1},
  pdfborder=0 0 1
}
\usepackage{natbib}
\setcitestyle{round}

\usepackage{fancyhdr}
\def\di{\mathrm{d}}
\def\xx{\mathrm{x}}
\def\Ry{\mathrm{Ry}}

\def\qq{\mathcal{Q}}
\def\opt{\mathrm{opt}}
\def\zz{\tilde{z}}
\def\kB{k_\mathrm{B}}
\def\atm{\mathrm{atm}}
\def\trfn{\Psi(z_\atm/r)}
\def\ar{a_\mathrm{r}}
\def\Cc{C'}
\def\dqc{\di q'}
\def\nc{\nabla'}
\def\nr{\nabla_\mathrm{r}}
\def\na{\nabla_\mathrm{ad}}
\def\A{\mathcal{A}}
\def\F{\mathcal{F}}
\def\leps{\lambda_\varepsilon}
\def\neps{\nu_\varepsilon}

\newcommand{\atpoint}[2]{\left.#1\right|_{#2}}
\sloppypar

\begin{document}
\baselineskip 21pt

\title{\bf Vertical Convection in Turbulent Accretion Disks and\\ Light Curves of the A0620--00 1975 Outburst }

\author{\bf K.~L.~Malanchev\affilmark{1,2*}, N.~I.~Shakura\affilmark{1}}

\affil{
{$^1$\it Sternberg Astronomical Institute, M.~V.~Lomonosov Moscow State University, Moscow, Russia }\\ 
{$^2$\it Faculty of Physics, M.~V.~Lomonosov Moscow State University, Moscow, Russia }}

\vspace{2mm}

\sloppypar 
\vspace{2mm}
\noindent
We present a model of the non-stationary $\alpha$-disk with account for the irradiation and the vertical convection in the outer accretion disk where hydrogen is partially ionized.
We include the viscous energy generation in the mix-length convection equations in accretion disks.
The optical and X-ray light curves of X-ray nova A0620--00 are investigated in terms of this model.
The turbulent viscosity parameter of the accretion disk is estimated, $\alpha = 0.5 \div 0.6$, which is necessary to explain the luminosity decay rate on the descending branch of the X-ray light curve for the A0620--00 1975 outburst.
The secondary luminosity maximum on the light curves is explained by assuming an additional injection of matter into the accretion disk from the optical companion.


\noindent
{\bf Astronomy Letters,} Vol. 41, No. 12, 2015. 

\noindent
{\bf DOI:} 10.1134/S1063773715120087

\vfill
\noindent\rule{8cm}{1pt}\\
{$^*$ E-mail: $<$malanchev@sai.msu.ru$>$}

\clearpage

\section{Introduction}
X-ray novae are close binary systems with a relativistic object (a black hole or a neutron star) and a low-mass Roche-lobe-filling star (see, e.g. \citet*{cherepashchuk2000}; \citet{postnov_yungelson2014}).
The time evolution of an accretion disk may be considered in terms of a linear problem, where the kinematic viscosity in the disk does not depend on its surface density $\Sigma_0$.
An analytic solution for the disk evolution in this case was obtained by \citet{lipunova2015}.
The standard disk accretion model developed more than forty years ago \citep{shakura1972,shakura_sunyaev1973} implies that the kinematic viscosity depends on the surface density, and, hence, the accretion disk evolution problem is nonlinear.
For an $\alpha$-disk, analytic solutions were found for various stages of accretion disk evolution \citep{lyubarskij_shakura1987,lipunova_shakura2000}.

Analytic solutions of the nonlinear disk accretion problem are possible only in the case where an analytic relation exists between the accretion disk viscosity and surface density (for example, using the solution of the vertical disk structure equations; see \citet{ketsaris_shakura1998,lipunova_shakura2000}).
In the remaining cases, the vertical structure equations and the viscous accretion disk evolution equations should be solved numerically.
The numerical simulations of light curves by taking into account the thermal instability, convection, and self-irradiation in the case of dwarf novae, where matter is accreted onto a black hole \citep[e.g.][]{hameury_etal1998}, and X-ray novae \citep[e.g.][]{dubus_etal2001,wisniewicz_etal2015} were considered in a number of papers.
The main difference of the present work from the previous papers is a study of an impact of the turbulence on the vertical convection in the outer disk.
We describe the outer disk vertical structure using the mixing length theory in the Appendix.

The X-ray nova A0620--00 1975 outburst is one of the brightest X-ray novae and has been studied thoroughly.
The X-ray curve of this outburst rises on scales of several days and then decays quasiexponentially on scales of tens of days.
In this paper, we consider only the descending branches of the Xray and optical light curves for the X-ray nova A0620--00.

The X-ray and optical light curves of X-ray nova A0620--00 within the first 50 days after its peak were jointly modeled in \citet{suleimanov_etal2008}, where $\alpha \gtrsim 0.5$ were obtained.
In this paper, we model the light curves of this nova within the first 120 days after its peak.
A characteristic feature of the light curves for A0620--00 is the secondary brightness peak observed in both X-ray and optical bands approximately on the 55th day after the primary maximum.
The nature of this secondary maximum remains unclear and can be explained by a host of various physical processes.
One of the explanations can be the radiative heating of the hitherto cold outer disk \citep{king_ritter1998,ertan_alpar2002}.
Another suggestion is the evaporation of matter in the central hot disk regions and their subsequent filling with material \citep{cannizzo2000}.
Lipunova and Shakura in their \citeyear{lipunova_shakura2001} and \citeyear{lipunova_shakura2003} papers associate the appearance of a secondary peak with the onset of convection in the outer accretion disk, which can effectively increase the viscous friction.
In this paper, we use the assumption about an additional injection of mass into the disk from the optical companion.
Such an ejection can be explained both by the irradiation of the optical star by X-rays from the accretion disk and by internal instabilities in its convective envelope.

\section{Viscous evolution of an accretion disk}
We will consider a thin accretion disk symmetric relative to the vertical axis and the midplane in which the velocities of the accreting material depend only on its distance $r$ to the disk center.
In such a disk, the continuity equation integrated along the vertical axis is
\begin{equation}
	\frac{\partial \Sigma_0}{\partial t} = - \frac1{r} \frac{\partial}{\partial r} (\Sigma_0 v_r r),
\label{eq.cont}
\end{equation}
where $\Sigma_0 = \int_{-\infty}^{\infty}{\rho \di z}$ is the surface density of the ring
of material at radius $r$, and $v_r$ is the radial velocity of the material in the disk.

The angular momentum transport equation can be written as
\begin{equation}
	\Sigma_0 v_r \frac{\partial (\omega r^2)}{\partial r} = \frac1{r} \frac{\partial}{\partial r}(W_{r\varphi}r^2),
\label{eq.ang_mom_transfer}
\end{equation}
where $\omega = \sqrt{G M_\xx/r^3}$ is the angular velocity of the accreting material in the disk, $G$ is the gravitation constant, $M_\mathrm{x}$ is a black hole mass, and $W_{r\varphi} = \int_{-\infty}^{\infty}{w_{r\varphi} \di z}$ is the $r\phi$-component of the viscous stress tensor integrated along the vertical axis.
Here, we neglect the tidal torques by assuming them to be negligible in the entire disk, except for its outermost ring \citep{ichikawa_osaki1994}.

Let us introduce the viscous torque acting between adjacent layers $F$:
\begin{equation}
	F = - 2\pi W_{r\varphi} r^2.
\label{eq.vis_torque}
\end{equation}

Substituting \eqref{eq.ang_mom_transfer} and \eqref{eq.vis_torque} into \eqref{eq.cont} and using the specific angular momentum $h = \sqrt{G M_\xx r}$ as a new radial coordinate, we will obtain a diffusion-type differential equation:
\begin{equation}
	\frac{\partial \Sigma_0}{\partial t} = \frac1{4\pi} \frac{(GM_\xx)^2}{h^3} \frac{\partial^2 F}{\partial h^2}.
\label{eq.diffusion}
\end{equation}
The initial and boundary conditions should be specified to solve this equation.
The material from the inner disk edge falls without viscosity, moving in the region of unstable circular orbits toward the black-hole event horizon in the dynamical time.
On this basis, we will set the following inner boundary condition on the viscous torque:
\begin{equation}
	\atpoint{F}{r=r_\mathrm{in}} = 0,
\label{eq.in_bound_cond}
\end{equation}
where \citep*{book_black-hole_accretion_disks}:
\begin{eqnarray}
	&r_\mathrm{in} &=\displaystyle \frac{GM_\xx}{c^2} (3 + Z_2 - \sqrt{(3-Z_1)(3 + Z_1 + 2Z_2)}) \label{eq.r_in}\\
    && \notag \textrm{ is the radius of the innermost stable circular orbit} ,\\
	&Z_1 &= 1 + \sqrt[3]{ 1 - a^2} (\sqrt[3]{1+a} + \sqrt[3]{1-a}),\\
	&Z_2 &= \sqrt{ 3a^2 + Z_1^2 },\\
	&a &\textrm{ is the Kerr parameter of the black hole},\\
	&c &\textrm{ is the speed of light}.
\end{eqnarray}

Note that $2 \pi \Sigma_0(r) v_r(r) r$ is equal to the accretion rate $\dot{M}(r,t)$.
Expressing the left-hand side of Eq.~\eqref{eq.ang_mom_transfer} in terms of the accretion rate, we will obtain the relation between the accretion rate and the viscous torque:
\begin{equation}
	\dot{M}(h) = - \frac{\partial F}{\partial h}
\label{eq.mdot}
\end{equation}
At the outer disk edge, where there are no inflow and outflow of matter,
\begin{equation}
	\left.\frac{\partial F}{\partial h}\right|_{h_\mathrm{out}} = 0,
\label{eq.out_bound_cond}
\end{equation}
where $h_\mathrm{out} = \sqrt{GM_\xx r_\mathrm{out}}$ is the specific angular momentum at the outer disk radius.
The outer disk radius is determined by the tidal radius, which we assume to be 80\% of the effective Roche lobe radius for a black hole \citep{eggleton1983,paczynski1977,suleimanov_etal2008}:
\begin{equation}
	r_\mathrm{out} = 0.8 \tilde{a} \frac{ 0.49 (M_\xx/M_\opt)^{2/3} }{ 0.6 (M_\xx/M_\opt)^{2/3} + \ln{(1 + (M_\xx/M_\opt)^{1/3})} },
\label{eq.r_out}
\end{equation}
where $\tilde{a}$ is the semimajor axis of the binary system, and $M_\opt$ is the mass of the optical companion.

We choose the initial condition in a form satisfying the boundary conditions:
\begin{equation}
	\atpoint{F(h)}{t=0} = \frac2{\pi} (h_\mathrm{out}-h_\mathrm{in}) \dot{M}_0 \times \sin{\left( \frac{\pi}{2} \frac{h-h_\mathrm{in}}{h_\mathrm{out}-h_\mathrm{in}} \right)},
\label{eq.init_cond}
\end{equation}
where $\dot{M}_0 \equiv \dot{M}(h_\mathrm{in})\left.\right|_{t=0}$ is the accretion rate onto the black hole at the peak of X-ray luminosity.

The relation between the two unknown function $\Sigma_0(h,t)$ and $F(h,t)$ should be established to solve the system of equations \eqref{eq.in_bound_cond}, \eqref{eq.out_bound_cond} and \eqref{eq.init_cond}; it can be found from the solution of the vertical structure equation for the accretion disk (see the next Section). 
This system was solved numerically using the implicit method that guaranteed the stability of the solution.
The solution was performed on a logarithmic spatial grid in $h$ with a crowding near $h_\mathrm{out}$ consisting of 400 grid points.
The time step was constant and equal to 0.2 day.
This exceeds the thermal time in the outer disk and allows us to disregard the effects associated with the fact that the radiation and matter in the disk are nonequilibrium ones during its abrupt cooling.

\section{Vertical structure}
We perform an independent calculation of the vertical disk structure at various radii.
The calculation is performed in the approximation of hydrostatic equilibrium and the Eddington approximation for radiative transfer.
For the convenience
of our calculations, the vertical coordinate $\zz$ in the disk is measured from the photosphere toward the disk midplane.
In view of the disk symmetry relative to the midplane, the vertical structure equations are
solved only between the points with vertical coordinates $\zz = 0$ (optical disk photosphere) and $\zz = z_0$ (disk midplane).
The value of $z_0$ corresponds to the half-thickness of a standard accretion disk.

\subsection{Hydrostatic Equilibrium Equation}
The hydrostatic equilibrium equation is
\begin{equation}
	\frac{\di P}{\di \zz} = \rho g_z,
\label{eq.hydrostat_gz}
\end{equation}
where $P$ is the pressure, $\rho$ is the density, $g_z = G M_x / r^2 \times (z_0-\zz)/r$ is the vertical component of gravity.
We neglect the radiation pressure in our calculations.
It is convenient to express $g_z$ in \eqref{eq.hydrostat_gz} in
terms of the angular velocity of the accreting material $\omega = \sqrt{G M_\xx / r^3}$:
\begin{equation}
	\frac{\di P}{\di \zz} = \rho \omega^2 (z_0 - \zz).
\label{eq.hydrostat}
\end{equation}
Let us write the derivative of the pressure with respect to the optical depth $\tau$ measured along the vertical $\zz$ axis:
\begin{equation}
	\frac{\di P}{\di \tau} = \frac{\omega^2 (z_0-\zz)}{\varkappa},
\label{eq.dp_dtao}
\end{equation}
where $\varkappa$ is the Rosseland opacity.
We use the tabulated opacities from the OPAL project \citep{opal} and \citet{ferguson_etal2005} for solar chemical composition \citep{solar_composition2009}.
The pressure at the level of the accretion disk photosphere where the optical depth $\tau = 2/3$ is estimated to be
\begin{equation}
	\atpoint{P}{\zz=0} = \frac23 \frac{\omega^2 z_0}{\varkappa}.
\label{eq.P_0}
\end{equation}

\subsection{Equation for the Surface Density}
Neglecting the surface density of the disk layers above the optical photosphere, we have
\begin{equation}
	\frac{\partial \Sigma}{\partial \zz} = 2 \rho,
\label{eq.dSigma_dz}
\end{equation}
where $\Sigma$ is twice the surface density of the disk layer between its photosphere and the current vertical coordinate $\zz$.
The boundary condition for Eq. \eqref{eq.dSigma_dz} can be written as
\begin{equation}
	\atpoint{\Sigma}{\zz=0} = 0.
\label{eq.Sigma_0}
\end{equation}

By integrating system \eqref{eq.dSigma_dz} and \eqref{eq.Sigma_0}, we will obtain
the total surface density $\Sigma_0$:
\begin{equation}
	\Sigma_0 = \atpoint{\Sigma}{\zz=z_0}.
\end{equation}
Note that the latter relation is not a boundary condition, because $\Sigma_0$ is assumed to be unknown when integrating the vertical structure equations.

\subsection{Energy Equation}
In the $\alpha$-disk model, the generation of energy per unit time per unit volume by viscous forces is $w_{r\varphi} r \di \omega/\di r = \frac32 \alpha P \omega$ \citep{shakura_sunyaev1973}.
In the outer accretion disk regions, the rate of energy release through the thermalization of X-ray radiation $\epsilon_\xx(\zz)$ arrived from its inner parts and incident on the
surface is a significant source of energy:
\begin{equation}
	\frac{\di Q}{\di \zz} = - \frac32 \alpha P \omega - \epsilon_\xx(\zz).
\label{eq.energy_gen}
\end{equation}

Given \eqref{eq.vis_torque}, the total energy flux being released in the disk through turbulent viscosity $Q_\mathrm{vis}$ is
\begin{equation}
	Q_\mathrm{vis} = \int_{-\infty}^{+\infty}{w_{r\varphi} r \frac{\di \omega}{\di r} \di z} = -\frac32 W_{r\varphi} \omega = \frac3{8\pi} \frac{F \omega}{r^2}.
\label{eq.Q_vis}
\end{equation}

\begin{figure}[t]
	\includegraphics[scale=1.]{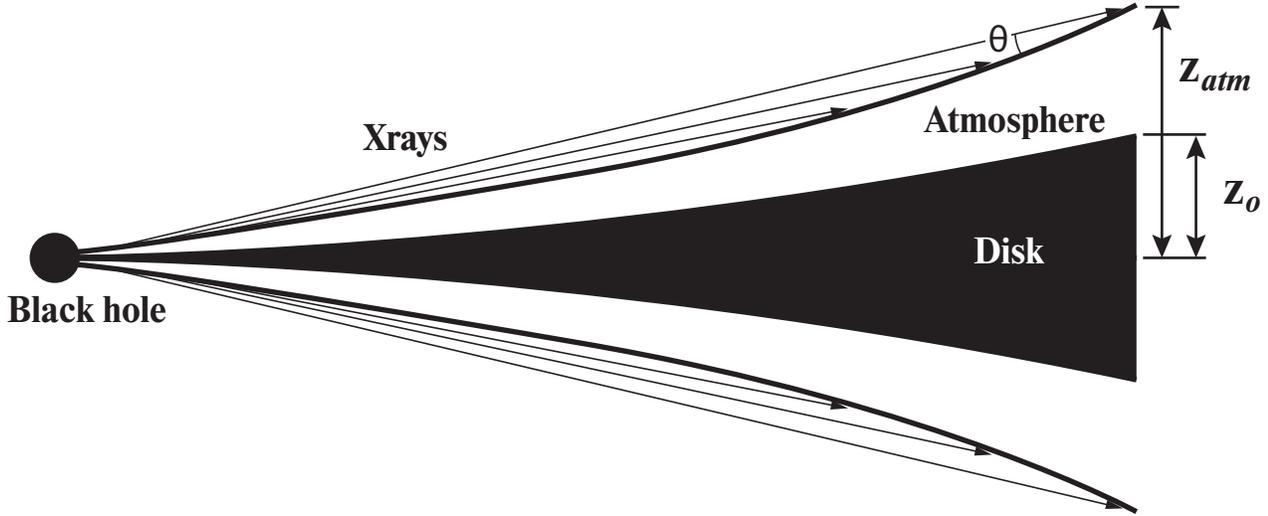}
    \caption{ \small \label{fig.disk} Schematic view of an accretion disk with an atmosphere and X-ray irradiation.}
\end{figure}

To calculate the flux of the X-ray radiation, we used a two-component disk model (see Fig. \ref{fig.disk}).
In this model, there is a hot atmosphere above the optical disk photosphere that scatters and absorbs some of the X-ray photons.
The disk below the optical photosphere is assumed to be cold relative to the X-ray radiation and the absorption in it dominates over the scattering.
The disk under consideration is optically thick to the X-ray irradiation, and the energy release near its midplane is determined exclusively by viscosity.
This allows the model of a semi-infinite plane-parallel layer to be used in writing the X-ray radiation transfer equations.

Let us write the X-ray flux $Q_\xx$ incident on the hot atmosphere by assuming its boundary to be located at height $z_\atm$ above the disk midplane (i.~e. $\zz = z_0 - z_\atm$):
\begin{equation}
	\atpoint{Q_\xx}{\zz = z_0 - z_\atm} = \frac{L_\xx}{4\pi r^2}\, \trfn\, \sin{\theta},
\label{eq.Q_x_atm}
\end{equation}
where $L_\xx = \tilde{\eta} \dot{M}_\mathrm{in}$ is the X-ray luminosity of the disk, $\tilde{\eta}$ is the accretion efficiency, $\trfn$ is the transfer function of the inner disk for a remote observer, and $\theta$ is the angle between the X-ray photon propagation direction and the surface of the atmosphere.
We used the transfer function from \citet{suleimanov_etal2008}.
For a Schwarzschild black hole and a relation $z_\mathrm{atm} / r = 0.1$ that is typical for the outer disk, the value of $\Psi(z_\mathrm{atm} / r)$ is about 0.35.

The angle $\theta$ in the thin-disk approximation can be written as
\begin{equation}
	\theta \simeq \frac{\di z_\atm}{\di r} - \frac{z_\atm}{r}.
\label{eq.theta_desc}
\end{equation}
According to numerical models for the atmospheres of accretion disks in X-ray binaries \citep[e.g.][]{jimenez-garate_etal2002}, to a first approximation, it can be assumed that $z_\atm = k_z z_0$ in the outer disk, where the irradiation is significant.
The expression for $\theta$ will then take the form
\begin{equation}
	\theta = k_z \left( \frac{\di z_0}{\di r} - \frac{z_0}{r} \right) = k_z \frac{z_0}{r} \left( \frac{\di \ln{z_0}}{\di \ln{r}} - 1 \right).
\label{eq.theta}
\end{equation}
Note that, formally, $\theta$ can take on negative values in the outer disk when convection sets in.
In our calculations, we proceeded from the fact that the change in the shape of the photosphere at the onset of convection in the disk did not affect the shape of the surface of the atmosphere and used a constant value for $\di \ln{z_0} / \di \ln{r} = 0.06$, which we obtained for the outer disk at temperatures above 10~000~K.

According to \citet{mescheryakov_etal2011a}, the X-ray flux at the level of the optical disk photosphere can be written as
\begin{equation}
	\atpoint{Q_\xx}{\zz = 0} = \lambda^\atm\, e^{-k^\atm \tau_\xx^\atm}\, \atpoint{Q_\xx}{\zz = z_0 - z_\atm}
\label{eq.Q_x_0}
\end{equation}
where $\lambda^\atm \equiv \varkappa_\mathrm{T} / \varkappa_\xx^\atm$ is is the ratio of the Thomson opacity to the total opacity for X-ray photons in the atmosphere, $\tau_\xx^\atm$ is the optical depth of the atmosphere along the $\zz$ axis for X-rays, and $k^\atm \equiv \sqrt{3(1-\lambda^\atm)}$.

Let us designate the attenuation of the X-ray flux in the atmosphere as $(1 - \tilde{A})$:
\begin{equation}
	1 - \tilde{A} \equiv \lambda^\atm\, e^{-k^\atm \tau_\xx^\atm}.
\label{eq.albedo_atm}
\end{equation}
Using this designation, we will write the X-ray flux under the optical photosphere \citep{mescheryakov_etal2011a} as
\begin{equation}
	\atpoint{Q_\xx(\zz)}{\zz \geq 0} = (1 - \tilde{A})\, \frac{L_\xx}{4\pi r^2}\, \trfn\, \sin{\theta}\, e^{- k_\mathrm{d}\varkappa_\xx \Sigma(\zz)},
\label{eq.Q_x}
\end{equation}
where $\varkappa_\xx$ is the opacity for X-rays in a cold medium, was chosen to be 5.7~cm$^2$~g$^{-1}$, corresponding to the opacity for photons with an energy of 3 keV in cold matter of solar chemical composition \citep{balucinska-church_mccammon1992}, $\lambda_\mathrm{d} = 0$  (the Thomson opacity is much smaller than the opacity for absorption), $k_\mathrm{d} = \sqrt{3(1-\lambda_{d})} = \sqrt{3}$.

Having differentiated \eqref{eq.Q_x} with respect to $\zz$, we obtain the final expression for the rate of energy release per unit volume $\epsilon_\xx$ through the thermalization of X-ray radiation:
\begin{equation}
	\epsilon_\xx(\zz) = k\, \frac{L_\xx}{4\pi r^2}\, \trfn\, \frac{z_0}{r}\, \left( \frac{\di \ln{z_0}}{\di \ln{r}} - 1 \right)\, k_\mathrm{d} \varkappa_\xx \rho\, e^{- k_\mathrm{d}\varkappa_\xx \Sigma(\zz)},
\label{eq.epsilon_x}
\end{equation}
where $k \equiv (1 - \tilde{A}) k_z$ is the unknown coefficient associated with the atmospheric structure, which can be determined by modeling the optical light curve of the X-ray nova.

Using \eqref{eq.Q_vis} и \eqref{eq.Q_x_0}, we will write the total flux of thermal radiation through the disk photosphere as
\begin{equation}
	\atpoint{Q}{\zz=0} = \frac3{8\pi} \frac{F\omega}{r^2} + k\, \frac{L_\xx}{4 \pi r^2}\, \trfn\, \frac{z_0}{r}\, \left( \frac{\di \ln{z_0}}{\di \ln{r}} - 1 \right).
\label{eq.Q_0}
\end{equation}

In view of the disk symmetry relative to the midplane, the boundary condition for the energy flux in the plane of disk symmetry is
\begin{equation}
	\atpoint{Q}{\zz=z_0} = 0.
\label{eq.Q_z0}
\end{equation}

\subsection{Energy Transfer Equation}
The equation of energy transfer by radiation $Q_\mathrm{rad}$ in the diffusion approximation can be written as
\begin{equation}
	\frac{c}{3 \varkappa \rho} \frac{\di \ar T^4}{\di \zz} = Q_\mathrm{rad},
\label{eq.rad_transfer}
\end{equation}
where $\ar$ is the radiation constant, $T$ is the temperature of the accreting matter.

At a temperature below 10~000~K, hydrogen becomes partially ionized and the opacity increase sharply, which gives rise to a sharp increase in the vertical temperature gradient and convection.
An expression for $\di T / \di \zz$ n this case was derived in the Appendix~\eqref{eq.dTdz_conv} in terms of themixing-length theory.

Thus, the vertical temperature gradient will be defined by Eq.~\eqref{eq.dTdz_conv} when the criterion for the onset of convection~\eqref{eq.conv_ineq_ln} is met and by Eq.~\eqref{eq.rad_transfer} in the opposite case:
\begin{equation}
\frac{\di T}{\di \zz} = \left\{ \begin{split} &\frac{3 \varkappa \rho}{4 \ar c T^3} Q,\, \nr < \na; \\ & \frac{g_z \rho T}{P} \nabla,\, \nr \geq \na \end{split}. \right.
\label{eq.dTdz}
\end{equation}

The boundary condition for the temperature at the level of the disk photosphere is determined by the effective temperature:
\begin{equation}
	\atpoint{T}{\zz=0} = \sqrt[4]{ \frac{4 \atpoint{Q}{\zz=0}}{\ar c} }.
\label{eq.T_0}
\end{equation}

\subsection{Solving the System of Vertical Structure Equations}
Equations \eqref{eq.hydrostat}, \eqref{eq.dSigma_dz}, \eqref{eq.energy_gen}, and \eqref{eq.dTdz} together with the boundary conditions \eqref{eq.P_0}, \eqref{eq.Sigma_0}, \eqref{eq.Q_0}, \eqref{eq.Q_z0}, and \eqref{eq.T_0} form a system of four linear differential equations with five boundary conditions (four are specified on the disk surface and one is specified in the plane of symmetry) and one unknown, $z_0$.
The disk halfthickness $z_0$ is chosen in such a way that when this
system of equations is integrated from point $\zz = 0$,
at which the boundary conditions are specified for all
unknown functions $P(\zz)$, $\Sigma(\zz)$, $Q(\zz)$, and $T(\zz)$, to
point $\zz = z_0$, the boundary condition $\atpoint{Q}{z=z_0}=0$ is met.

As a result, for given radius, viscous torque, and X-ray flux incident on the disk photosphere, we can obtain the vertical distributions of all thermodynamic quantities and the value of the disk surface density.
The derived surface density profile for a set of accretion disk rings can be used to solve the viscous disk evolution Eq.~\eqref{eq.diffusion}.
To optimize the time it takes to model the light curves, we used the table of integral disk parameters ($\Sigma_0$, $z_0$, etc.) as a function of three parameters: the specific angular momentum $h$, the viscous torque $F$, and the ratio of the total flux from the disk surface to the viscous flux $f_Q = \atpoint{Q}{\zz=0} / Q_\mathrm{vis}$.
The table was compiled for the logarithms of $h$, $F$, and $f_Q$ and had the dimensions ($500 \times 500 \times 100$).
The table was filled with the values, as necessary, when numerically solving Eg.~\eqref{eq.diffusion}; the final filling with the elements does not exceed 1\% of the number of cells in the table.

\section{Modeling the light curves of X-ray nova A0620--00}
To test the described numerical model, we chose the well-known X-ray nova A0620--00.
The advantage of this system is that the orbital parameters needed in our modeling are well known for it and there are both X-ray and optical observations of its 1975 outburst \citep{cherepashchuk2000,chen_etal1997}.

The observed light curves of this nova contain a secondary maximum on their exponential branches.
To explain this phenomenon, we use the hypothesis about an additional injection of matter from the secondary component into the accretion disk \citep{lipunova2015}.
For this purpose, we instantaneously increase the surface density at radii $r \geq 0.8 r_\mathrm{out}$ on the 43th day after the primary luminosity maximum.
When the effective temperature of the outer accretion disk drops below 10~000~K, it becomes important to take into account the thermal instability.
The latter enables a rapid transition of the disk ring to the cold state, whereby the disk hydrogen becomes neutral \citep[e.g.][]{hoshi1979,meyer_meyer-hofmeister1981,smak1982,cannizzo_etal1988}.
By the time of the transition to the cold state, the effective temperature is $6000\div7000$~K and the mass fraction of matter involved in convection at the radius under consideration is more than 80\%.
During our numerical simulations, we assumed that the outer radius of the hot disk was shifted inward together with the thermal instability zone and that the accretion in the outer cold disk stopped completely \citep{lipunova_shakura2003}.

\begin{figure}[t!]
\centering
	\includegraphics[scale=1.]{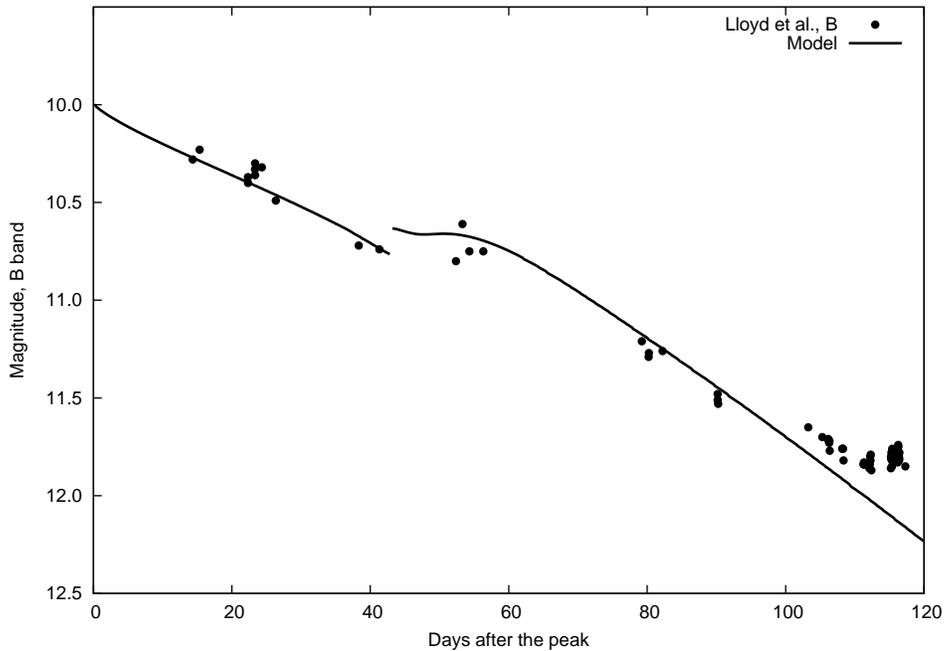}
    \caption{ \small \label{fig.b_band} Optical light curve of X-ray nova A0620--00. The optical observations \citep{lloyd_etal1977} are shown; the solid line indicates our model for the black hole mass of~$12 M_\odot$.}
\end{figure}

\begin{figure}[t!]
\centering
	\includegraphics[scale=1.]{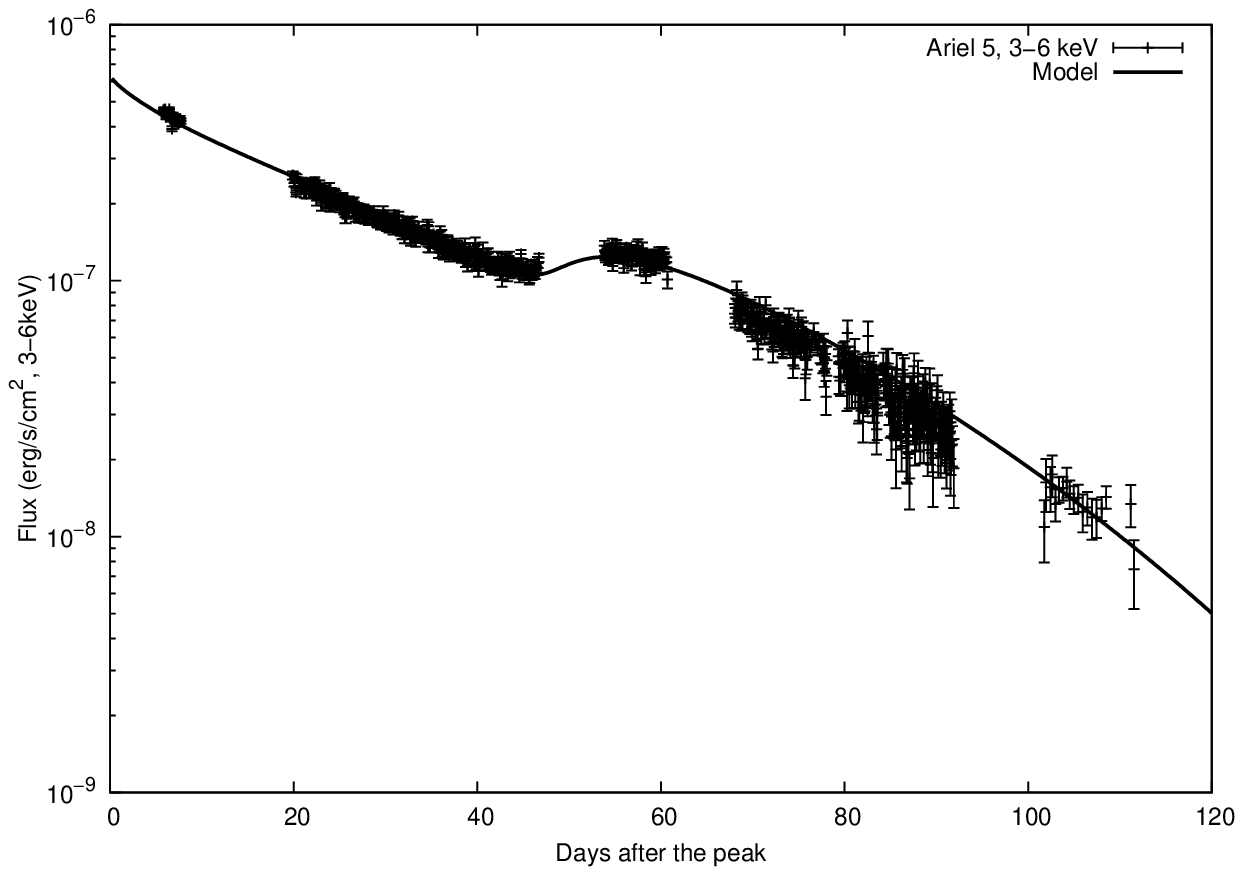}
    \caption{ \small \label{fig.x_ray} Soft X-ray (3---6~keV) light curve of X-ray nova A0620--00. The Ariel 5 observations \citep{kaluzienski_etal1977} are shown; the solid line indicates our model for the black hole mass of~$12 M_\odot$. }
\end{figure}

In our modeling, we used the following parameters of the binary system: the mass of the optical companion $M_\opt = 0.4 M_\odot$, the orbital period $P = 0.323$ day, and the Kerr parameter of the black hole $a = 0.2$ \citep{chen_etal1997,cherepashchuk2000,gou_etal2010}.
We performed our modeling for two black hole masses: $12 M_\odot$ \citep{gelino_etal2001} and $6.6 M_\odot$ \citep{cantrell_etal2010}; orbital inclinations of $38^\circ$ and $51^\circ$, respectively, at the system’s known mass function of $2.7 M_\odot$ correspond to these parameters.
The modeling results for the black hole mass of $12 M_\odot$ are presented in Figs. \ref{fig.b_band} and \ref{fig.x_ray}; the parameters of both models are given in Table \ref{table.parameters}.
To achieve agreement between the model and observations, the values for two model parameters were chosen in the segment of the X-ray light curve before the secondary peak: the turbulent viscosity parameter $\alpha$ and the accretion rate at maximum light $\dot{M}_0$ in units of $(L_\mathrm{Edd} / c^2)$, where $L_\mathrm{Edd}$ is the Eddington luminosity for a black hole of the corresponding mass.
To explain the secondary peak on the X-ray light curve, we chose the amount of matter added to the outer disk on the 43rd day after the light-curve maximum $\delta M$ in fractions of the then available mass in the disk.
At the same time, the optical light curve (Fig. \ref{fig.b_band}) exhibits an abrupt rise in the luminosity due to the assumption about an instantaneous increase in the mass.

\begin{table}[t!]
	\centering
	\begin{tabular}{l|l|l}
		\hline
		Parameter													& $M_\xx = 6.6 M_\odot$				& $M_\xx = 12 M_\odot$\\
		\hline
		Distance to binary system, kpc								& 0.85  & 1.0\\
		Turbulent viscosity parameter $\alpha$						& 0.5   & 0.6\\
		Initial accretion rate $\dot{M}_0$, $L_\mathrm{edd}/c^2$	& 1.8	& 1.9\\
		Ratio of added mass to disk mass $\delta M$ on 43th day     & 0.3	& 0.25\\		
		Atmospheric irradiation interception parameter $k$		    & 1.4	& 0.8\\
		\hline
	\end{tabular}
	\caption{ \small \label{table.parameters} Parameters of the models for two different black hole masses obtained for the light curves of the 1975 outburst of A0620--00 }
\end{table}

Once these parameters were established, we chose the parameter of the disk atmosphere k to explain the $B$ light curve.
Just as in \citet{suleimanov_etal2008} we used parameter $k_z \equiv z_\atm/z_0 = 2$ in modeling A0620--00.
Note that \citet{mescheryakov_etal2011b} derived the same relation between the thickness of the atmosphere and the standard disk thickness at the outer radius from observations of the low-mass X-ray binary GS 1826–238.
Using this value of $k_z$, we can obtain $\tilde{A} = 1 - k/k_z$, the fraction of the radiation reflected or absorbed in the disk atmosphere, which is 0.3 and 0.6 for the models with the black hole masses
of $6.6 M_\odot$ and $12 M_\odot$, respectively.
Assuming that the optical depth of the atmosphere for X-rays $\tau_\xx^\mathrm{atm}$ is greater than unity, we can place a constraint on the ratio of the scattering coefficient to the total absorption coefficient in the atmosphere from \eqref{eq.albedo_atm}: $\lambda_\mathrm{atm} > 0.96$ and $\lambda_\mathrm{atm} > 0.82$ for the black hole masses of $6.6 M_\odot$ and $12 M_\odot$, respectively.

The optical flux deficiency on the model light curve (Fig.~\ref{fig.b_band}) after the 100th day can be explained by the appearance of an additional source of optical radiation.
The cold parts of the disk irradiated by scattered X-ray radiation in the atmosphere, whose area increases with time but whose radiation is disregarded when constructing the light curve, can be such a source.
The X-ray-heated surface of the optical companion can be another possible source.
As the radius of the hot disk decreases, the ratio $z_0/r$ (Fig.~\ref{fig.disk}) and, as a consequence, the size of the shadow cast by the X-ray radiation from the outer disk decrease, increasing the area of the irradiated surface of the star.

\begin{figure}[t!]
\centering
	\includegraphics[scale=1.]{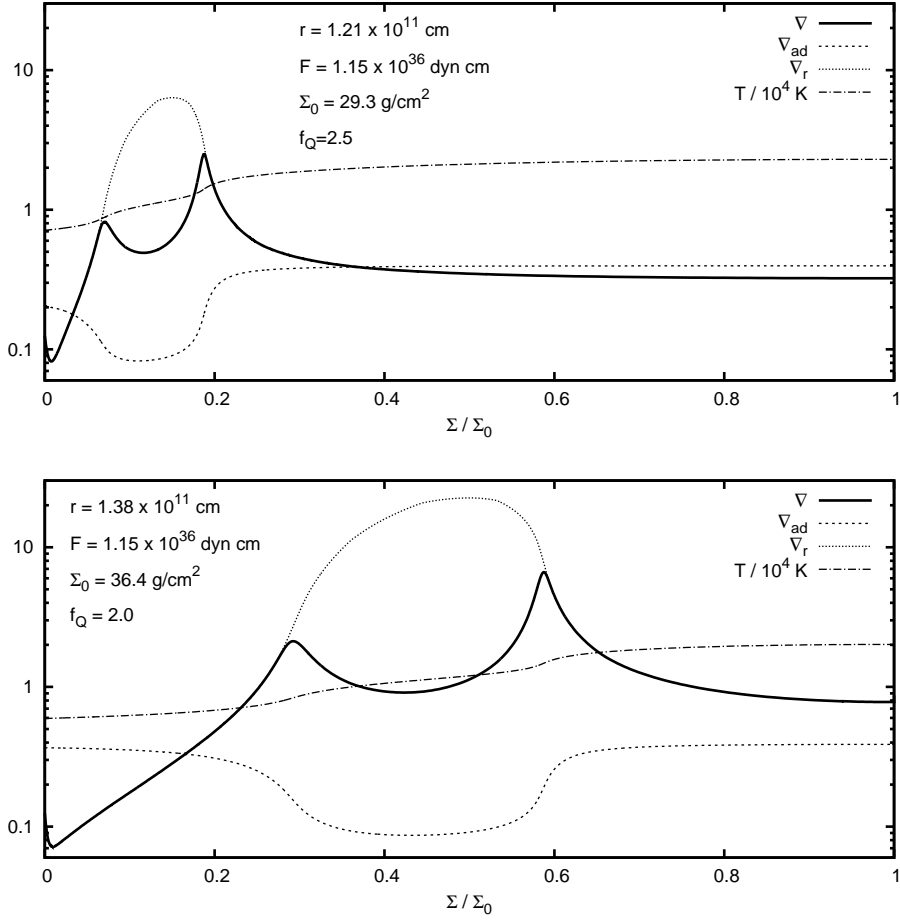}
    \caption{ \small \label{fig.vs} Vertical distributions of the temperature and three logarithmic gradients, the mean $\nabla$, the adiabatic $\na$, and the fictitious radiative $\nr$. Convection takes place under the condition $\nr > \na$. The mass coordinate $\Sigma / \Sigma_0$, which is zero at the level of the optical photosphere and one in the plane of symmetry, is along the horizontal axis. The distributions are presented for the model with the black hole mass of~$12 M_\odot$ on the 80th day after the light-curve maximum. The upper and lower panels correspond to radii of $1.21 \times 10^{11}$~cm and $1.38 \times 10^{11}$~cm (the outer radius of the hot disk), respectively.}
\end{figure}

\section{Conclusions}
The light curves of X-ray nova A0620--00 after its 1975 outburst were modeled in terms of the standard nonstationary $\alpha$-disk model.
The descending branches of the optical and X-ray light curves, including the secondary luminosity maximum, were investigated.
We showed that this secondary maximum could be explained by an additional injection of matter into the accretion disk from the donor star.
At the early outburst stage, the disk is in a hot state with its outer radius determined according to \citet{paczynski1977}.
After the secondary luminosity maximum, approximately on the 60th day after the primary maximum, a zone with partially ionized hydrogen appears in the outer disk, which is shifted toward the disk center with time.
A cold disk in which hydrogen is completely neutral and the accretion is very slow is formed behind the front of partially ionized hydrogen.

Vertical convection and irradiation in the outer accretion disk affect significantly its vertical structure.
Figure~\ref{fig.vs} shows the vertical distributions of the temperature and various logarithmic temperature gradients (see the Appendix) on the 80th day after the primary luminosity maximum for the model with a black hole mass of~$12 M_\odot$.
At this time, the outer radius of the hot disk is $1.38 \times 10^{11}$ cm, while the radius of the cold disk is $1.64 \times 10^{11}$~cm.
The lower graph in Fig.~\ref{fig.vs} shows the vertical structure at the outer radius of the hot disk, where hydrogen at the photospheric level is already neutral and convection affects $\simeq 80\%$ of the disk mass fraction.
The upper graph demonstrates the structure at a smaller radius, where the disk temperature is considerably higher and convection affects only about a third of the matter in the upper disk layers.
Nevertheless, there is no convection near the photosphere due to the decrease in the vertical temperature gradient $\di T / \di \zz$ associated with the thermalization of X-ray radiation incident on the photosphere.

\section*{Acknowledgments}
We thank G.V. Lipunova, A.V.Mescheryakov, and K.A. Postnov for fruitful discussions.
This work was supported by the Russian Science Foundation grant 14-12-00146.
We used the equipment funded by M.V.~Lomonosov Moscow State University Program of Development.

\appendix
\renewcommand{\theequation}{A.\arabic{equation}}
\section*{Appendix: The mixing-length theory of convection}

In this paper, the mixing-length theory of convection is used to calculate the convective zones of an
accretion disk.
Here, we will present only the key points of this theory by considering the aspects of its application for accretion disks in more detail (for a thorough presentation of the mixing-length theory of
convection, see, e.g., the book \citet*{book_cox_and_giulis}).

Generally, the energy flux through a surface of unit area along the $\zz$ axis is the sum of the radiative and convective fluxes:
\begin{equation}
	Q = Q_\mathrm{rad} + Q_\mathrm{conv}.
\label{eq.full_flux}
\end{equation}

An expression for the energy flux transferred by radiation can be derived in the diffusion approximation:
\begin{equation}
	Q_\mathrm{rad} = \frac{4 \ar c T^3}{3 \varkappa \rho} \frac{\di T}{\di \zz},
\label{eq.Q_rad}
\end{equation}
where $\ar$ is the radiation constant.

As long as the vertical temperature gradient $\di T / \di \zz$ is small, there is no convection and $Q_\mathrm{conv} = 0$.
According to the Schwarzschild criterion, convection sets in when the temperature gradient \eqref{eq.Q_rad} derived from Eq.~\eqref{eq.Q_rad} exceeds the adiabatic gradient:
\begin{equation}
	\left(\frac{\di T}{\di \zz}\right)_\mathrm{rad} > \left(\frac{\di T}{\di \zz}\right)_\mathrm{ad}.
\label{eq.conv_ineq}
\end{equation}
For the convenience of our consideration of convection, we will introduce four logarithmic derivatives of the temperature with respect to the pressure, also called the logarithmic temperature gradients: $\nabla \equiv \di \ln{T} / \di \ln{P}$, the logarithmic gradient calculated for the mean parameters of the medium at a given $\zz$; $\nc$, the mean gradient for the rising convective cells (i.e., those moving in a direction away from the disk midplane toward its photosphere); $\nr$, the fictitious radiative gradient, i.e., the gradient that would be without convection; $\na$, the adiabatic gradient that is a function of the thermodynamic state of matter.
Figure~\ref{fig.vs} shows the logarithmic gradients as functions of the mass coordinate $\Sigma/\Sigma_0$ at two different accretion disk radii on the 80th day after the primary luminosity maximum.

For a gas of pure hydrogen, the adiabatic gradient $\na$ can be written as \citep{book_cox_and_giulis}
\begin{equation}
	\na \equiv \left(\frac{\di \ln{T}}{\di \ln{P}}\right)_\mathrm{ad} = \frac{ 2 + i(1-i) \left( \frac52 + \frac{\Ry}{kT} \right) }{ 5 + i(1-i) \left( \frac52 + \frac{\Ry}{kT} \right)^2 },
\label{eq.dlnTdlnP}
\end{equation}
where $\Ry=2.180\times10^{-11}$~erg is the Rydberg constant, $i$ is the degree of hydrogen ionization (the ratio of the number of hydrogen ions to the total number of hydrogen ions and atoms) derived from the Saha equation \citep{book_landafshitz_stat1980}:
\begin{equation}
	i = \frac1{\sqrt{1 + P K_P}},
\label{eq.i}
\end{equation}
$K_P$ is the chemical equilibrium constant:
\begin{equation}
	K_P \equiv \left( \frac{2 \pi \hbar^2}{m_\mathrm{e}} \right)^{3/2} \left( \frac1{\kB T} \right)^{5/2} \exp{\left( \frac{\Ry}{\kB T} \right)}.
\label{eq.K_P}
\end{equation}

By the definition, the fictitious radiative gradient $\nr$ can be expressed from~\eqref{eq.Q_rad} by substituting the total flux $Q$ for the radiative one $Q_\mathrm{rad}$:
\begin{equation}
	\nr \equiv \frac{3 \varkappa \rho P}{4 \ar c T^4} \left( \frac{\di P}{\di \zz} \right)^{-1} Q.
\label{eq.nabla_r}
\end{equation}

The Schwarzschild condition (A.3) can now be rewritten using the logarithmic gradients:
\begin{equation}
	\nr > \na.
\label{eq.conv_ineq_ln}
\end{equation}

We will consider the case where the Schwarzschild condition holds.

Convective cells transfer heat due to a temperature difference between their contents and the ambient
medium. Having traversed some distance, a convective cell dissolves in the ambient medium and transfers its heat to the latter. Therefore, let us write the heat flux transferred by convection by assuming the pressure in the convective cells to coincide with the pressure in the medium:
\begin{equation}
	Q_\mathrm{conv} = \frac12 \rho v C_P \Lambda \left[ \frac{\di T}{\di \zz} - \left(\frac{\di T'}{\di \zz}\right) \right] = \frac12 \rho v C_P T \Lambda \frac{\di P}{\di \zz} (\nabla - \nc),
\label{eq.Q_conv}
\end{equation}
where $T'$ is the temperature in the rising convective cell, $\Lambda$ is the mixing length, i.e., the mean free path of a convective element, $v$ is the mean velocity of the convective elements,
\begin{eqnarray}
	\label{eq.Cp} C_P & = & \frac{\mathcal{R}}{\mu} \left[ \frac52 + \frac12 i(1-i) \left( \frac52 + \frac{Ry}{\kB T} \right) \right]\\
    \notag && \textrm{is the specific heat of the gas at constant pressure}, \\
	\label{eq.mu} \mu & = & \frac1{1+i} \textrm{ is the molar mass},
\end{eqnarray}
$\mathcal{R}$ is the universal gas constant.
The factor $1/2$ in \eqref{eq.Q_conv} appears because of the assumption that approximately half of the matter moves in the convective cells upward and the second half moves downward.

We assume that the mixing length $\Lambda = \beta z_0$, where $\beta$ is a constant in space and time, equal to $0.4$ in our calculations.

To find the mean velocity $v$, we will use the following considerations. Let us write the specific force acting on the rising element of matter:
\begin{equation}
	\rho \frac{\di^2 \delta z}{\di t^2} = \rho \frac{\di v}{\di t} = -g_z (\rho + \Delta\rho) + \frac{\di P}{\di \zz} = g_z \Delta\rho,
\label{eq.conv_element_force}
\end{equation}
where $\delta z$ is the distance traversed by the convective element after its formation, and $\Delta \rho$ is the difference between the density of the convective element and the mean density at a given $\zz$.
Neglecting the change in gravity with height, let us write the specific work done by this force to move the element through its free path $\Delta z$:
\begin{equation}
	W(\Delta z) = - \int_0^{\Delta z}{ g_z \Delta \rho(\delta z) d \delta z } = - \frac12 g_z \Delta \rho(\Delta z) \Delta z.
\label{eq.conv_element_work}
\end{equation}

Averaging over $\Delta z$ gives the mean work done to
move the element through the mean mixing length $\Lambda$:
\begin{equation}
	\overline{W}(\Lambda) = \frac1{4} W(\Lambda) = - \frac1{8} g_z \Delta \rho(\Lambda) \Lambda,
\label{eq.conv_element_average_work}
\end{equation}
where the factor $1/4$ is related to the variable cell velocity and variable vertical gravity $g_z$ \citep{book_cox_and_giulis}.
Assuming that half of this work is converted into kinetic energy and the second half goes into heating
due to the viscous friction of elements against one another, we will obtain
\begin{equation}
	\frac12 \rho v^2 = \frac12 \overline{W}(\Lambda) = - \frac1{16} g_z \Delta \rho(\Lambda) \Lambda.
\label{eq.conv_element_keenetic_energy}
\end{equation}
The energy release through the viscous friction of convective elements is much smaller than the turbulent one; therefore, it does not contribute to the energy equation\eqref{eq.energy_gen}.

Let us write the expression for $\Delta \ln{\rho}$ provided that the pressures in all convective elements at the same $\zz$ are equal:
\begin{equation}
	\Delta \ln{\rho} = \left[ \left( \frac{\partial \ln{\rho}}{\partial \ln{\mu}} \right)_{P,T} \left( \frac{\partial \ln{\mu}}{\partial \ln{T}} \right)_{P} + \left( \frac{\partial \ln{\rho}}{\partial \ln{T}} \right)_{\mu,P} \right] \Delta \ln{T}.
\label{eq.delta_ln_rho}
\end{equation}
Let us introduce a quantity $\qq$:
\begin{equation}
\qq \equiv 1 - \left( \frac{\partial \ln{\mu}}{\partial \ln{T}} \right)_P = 1 + \frac12 i(1-i) \left( \frac52 + \frac{Ry}{\kB T} \right).
\end{equation}
Assuming the role of the radiation pressure to be negligible, for an ideal gas of pure hydrogen we will obtain
\begin{equation}
	\Delta \ln{\rho} = - \left[ 1 - \left( \frac{\partial \ln{\mu}}{\partial \ln{T}} \right)_P \right] \Delta \ln{T} = - \qq \Delta \ln{T}.
\label{eq.delta_ln_rho_delta_ln_T}
\end{equation}

After the substitution of \eqref{eq.delta_ln_rho_delta_ln_T} into \eqref{eq.conv_element_keenetic_energy}, we will obtain the final expression for the mean velocity of the rising element $v$:
\begin{equation}
	v^2 = \frac{\qq g_z^2 \rho \Lambda^2}{8 P} (\nabla - \nc).
\label{eq.v2}
\end{equation}
The latter expression is valid only under the condition of subsonic motion of the convective cells, which holds in our calculations.

Substituting \eqref{eq.v2} into \eqref{eq.Q_conv}, we will obtain the final expression for the energy flux transferred by convection $Q_\mathrm{conv}$ in terms of the mean temperature gradient $\nabla$, the temperature gradient in the rising element of matter $\nc$, and the quantities that are functions of the thermodynamic state of matter:
\begin{eqnarray}
	Q_\mathrm{conv} &=& \frac{C_P \qq^{1/2} \rho^{5/2} g_z^2 T \Lambda^2}{4\sqrt{2} P^{3/2}} (\nabla - \nc)^{3/2},\label{eq.Q_conv_final}\\
	Q &=& Q_\mathrm{rad} + Q_\mathrm{conv} = \frac{4 \ar c T^4}{3 \varkappa P} \nabla + \frac{C_P \qq^{1/2} \rho^{5/2} g_z^2 T \Lambda^2}{4\sqrt{2} P^{3/2}} (\nabla - \nc)^{3/2}.\label{eq.Q_full}
\end{eqnarray}
The three gradients $\nabla$, $\nc$, and $\nr$ can now be related by using the definition of $\nr$~\eqref{eq.nabla_r}:
\begin{equation}
	\nr = \nabla + \frac{3 C_P \varkappa \qq^{1/2} \rho^{5/2} g_z \Lambda^2}{16 \sqrt2 \ar c P^{1/2} T^3} (\nabla - \nc)^{3/2}.
\label{eq.nablas}
\end{equation}

To determine $\nabla$, it remains to find its relation to $\nc$.
Considering the question of the convection efficiency $\Gamma$, i.e., of what “excess” fraction of heat a convective element transfers in its lifetime with respect to the heat that it radiates in the same time, will be helpful for us.
The loss of heat through radiation by a convective element from a unit area per unit time is
\begin{equation}
	\frac{4 \ar c T^3}{3\varkappa\rho} \frac{\Delta T}{\Lambda/2},
\label{eq.conv_element_radiative_loss}
\end{equation}
where $\Delta T$ is the temperature difference between the surface of the element (the mean ambient temperature) and its center averaged over its lifetime, and $\Lambda/2$ corresponds to the distance at which this temperature changes.
This formula is valid for an optically thick convective element, i.e., the condition $\varkappa \rho \Lambda/2 \gg 1$ must hold.
Multiplying \eqref{eq.conv_element_radiative_loss} by the area of the element $\A$ and by its lifetime $\Lambda/v$, we will obtain the energy $\F \Lambda/v$ dissipated by the element in its lifetime:
\begin{equation}
	\F \frac{\Lambda}{v} = \frac{4 \ar c}{3} \frac{T^3}{\varkappa \rho} \frac{\Delta T}{\Lambda/2} \frac{\Lambda}{v} \A.
\label{eq.conv_element_radiative_loss_lifetime}
\end{equation}

The “excess” of heat transferred by the convective element is $C_P \rho \Delta T_{max}V$, where $V$ is the volume of the element, $\Delta T_{max}$ is the temperature difference between the element and the ambient medium at an instant before the element dissolution, which is assumed to be twice the mean difference $\Delta T$ introduced above.
Dividing this expression by \eqref{eq.conv_element_radiative_loss_lifetime} and substituting the mean velocity $v$ from \eqref{eq.v2}, we will obtain an expression for the convection efficiency $\Gamma$:
\begin{equation}
	\Gamma = \frac3{16 \sqrt2 a_0} \frac{C_P \varkappa g_z \qq^{1/2} \rho^{5/2} \Lambda^2}{\ar c T^3P^{1/2}} (\nabla - \nc)^{1/2},
\label{eq.Gamma}
\end{equation}
where $1 / (2 a_0) \equiv V / (\A \Lambda)$ and is $1/6$ for a spherical
element and $1$ for a cubic one.
We used $a_0 = 9/4$ for which $V / \A = 2/9\Lambda$ \citep{book_cox_and_giulis}.
Comparing the derived expression with \eqref{eq.nablas}, we obtain
\begin{equation}
	\nr = \nabla + a_0 \Gamma (\nabla - \nc).
\label{eq.nabla_r_final}
\end{equation}
Let us express the logarithmic gradient $\nc$ in terms of $\na$ and the heat capacity in the rising convective cell $\Cc$ \citep{book_cox_and_giulis}:
\begin{equation}
	\nc = \na \frac1{1 - \Cc/C_P}.
\label{eq.C_conv_element}
\end{equation}
The heat capacity $\Cc \equiv \dqc / \di T$, where $q'$ is the specific (per unit mass) quantity of heat.
The rate of change of the specific quantity of heat $\dqc / \di t$ is equal to the difference between the “excess” specific rate of energy release in the convective cell $\Delta\varepsilon$ and the
specific radiation power carried away from the cell surface $\F$.

\begin{equation}
    \begin{split}
    	\frac{\Cc}{C_P} &= \left( \frac{\dqc}{\di T'} \right) \frac1{C_P} = \left( \frac{\dqc}{\di t} \right) \left( \frac{\di T'}{d t} \right)^{-1} \frac1{C_P} = \frac{(\Delta\varepsilon - \F)/(\rho V)}{C_P (dT' / dt)} = \\
        &= \frac{- \F [1 - \Delta\varepsilon / \F]/(\rho V)}{C_P [(d T' / dt) - (dT / dt)] + C_P(dT / dt)}.
    \end{split}
\label{eq.C_conv_over_CP}
\end{equation}
Note that $C_P \rho V [(d T' / dt) - (dT / dt)] / \F = \Gamma$, and, substituting the derived expression for $C/C_P$ into \eqref{eq.C_conv_element}, we obtain
\begin{equation}
	\frac{\Gamma}{1 - \eta} = \frac{\nabla - \nabla'}{\nabla' - \nabla_{ad}},
\label{eq.Gamma_over_one_minus_eta}
\end{equation}
where $\eta \equiv \Delta\varepsilon / \F$ is the ratio between the “excess” rate of energy release in the element and its losses through radiation.

In our case, there are two sources of energy release in the medium, turbulent viscosity and the thermalization of X-ray emission.
However, the thermalization of X-ray emission in cold matter does not depend on its thermodynamic state; hence, the “excess” energy release in the convective cells with be related only to the rate of energy generation by turbulent viscosity $\varepsilon_\mathrm{vis}$:
\begin{equation}
	\varepsilon_\mathrm{vis} = - \frac1{\rho} \frac{\di Q_\mathrm{vis}}{\di \zz } = \frac32 \frac{\alpha P \omega}{\rho}.
\label{eq.epsilon_vis}
\end{equation}
The following amount of “excess” energy will be released in the convective element in its lifetime:
\begin{equation}
	\Delta\varepsilon \rho V \frac{\Lambda}{v} = (\neps- \qq \leps)\, \ln{(\Delta T)}\, \varepsilon_\mathrm{vis}\, \rho V \frac{\Lambda}{v},
\label{eq.conv_element_extra_energy_lifetime}
\end{equation}
where for an ideal hydrogen gas
\begin{eqnarray}
	\label{eq.lambda_epsilon} \leps &\equiv& \left( \frac{\partial \ln{\varepsilon_\mathrm{vis}}}{\partial \ln{\rho}} \right)_T = - \frac{i(1-i)}{2 + i(1-i)},\\
	\label{eq.nu_epsilon} \neps &\equiv& \left( \frac{\partial \ln{\varepsilon_\mathrm{vis}}}{\partial \ln{T}} \right)_\rho = \frac{2 + i(1-i) \left(\frac52 + \frac{Ry}{\kB T}\right)}{2 + i(1-i)}.
\end{eqnarray}

We will obtain $\eta$ by dividing \eqref{eq.conv_element_extra_energy_lifetime} by \eqref{eq.conv_element_radiative_loss_lifetime}:
\begin{equation}
	\eta = \frac{3}{16 a_0} \frac{(\neps - \qq\leps) \varepsilon_\mathrm{vis} \varkappa \rho^2 \Lambda^2}{\ar c T^4}.
\end{equation}

For the convenience of our subsequent reasoning, let us introduce a quantity $\kappa$ equal to the ratio of the convective and radiative thermal conductivities of convective elements, which is a function of the thermodynamic state of matter and does not depend on its logarithmic gradients:
\begin{equation}
	\kappa \equiv \frac{\Gamma}{(\nabla - \nc)^{1/2}} = \frac3{16 \sqrt2 a_0} \frac{C_P \varkappa g \qq^{1/2} \rho^{5/2} \Lambda^2}{\ar c T^3 P^{1/2}}.
\label{eq.conv_element_condactivity_ratio}
\end{equation}

Equations~\eqref{eq.nabla_r_final},~(\ref{eq.Gamma_over_one_minus_eta}) and identity~\eqref{eq.conv_element_condactivity_ratio} form a system of three algebraic equations with three unknowns: $\nabla$, $\nc$, and $\Gamma$.
Let us introduce a new variable $\zeta \equiv (\nabla_r - \nabla) / (\nabla_r - \nabla_{ad})$.
After some transformations, the system will be reduced to an equation for this variable:
\begin{equation}
	(1 - \eta) \zeta^{1/3} + B \zeta^{2/3} + a_0 B^2 \zeta - a_0 B^2 = 0,
\label{eq.cubic}
\end{equation}
where $B \equiv [(\kappa^2/a_0)(\nr - \na)]^{1/3}$ and $\Gamma = B \zeta^{1/3}$.
This equation in the case of $\eta \geq 0$ under consideration has exactly one nonnegative solution for $\zeta$ (for a detailed consideration of the case of $\eta = 0$, see \citet{book_cox_and_giulis}).
However, this solution at $\eta > (1 + B)$ will have a value greater than one, which can correspond to a negative value of $\nabla$, but this case is not realized in our calculations.

Thus, we have obtained the final expression for the sought-for vertical temperature gradient in the case where convection sets in, which is used to solve the vertical structure equations for an accretion disk:
\begin{equation}
	\frac{\di T}{\di \zz} = \nabla \frac{\di P}{\di \zz} \frac{T}{P} = \frac{g_z \rho T}{P} \left[ \zeta \na + (1 - \zeta)\nr \right].
\label{eq.dTdz_conv}
\end{equation}

\bibliography{malanchev_shakura}

\end{document}